\documentclass{aa}
\usepackage{graphicx}
\usepackage{txfonts}
\usepackage{hyperref}
\usepackage{ae,aecompl}
\usepackage[T1]{fontenc}
\usepackage{amssymb,amsmath}
\usepackage{wasysym}
\usepackage{color}
\usepackage{epsfig}     
\usepackage{longtable}
\usepackage{pdflscape}
\usepackage{mathtools}
\usepackage{float}
\usepackage{footnote}
\usepackage{epstopdf}
\usepackage[colorinlistoftodos]{todonotes}

\DeclareRobustCommand{\ion}[2]{\textup{#1\,\textsc{\lowercase{#2}}}}

\begin{document}
\title{NGC~6705 a young $\alpha-$enhanced Open Cluster from OCCASO data}
\titlerunning{NGC 6705, a young $\alpha-$enhanced Open Cluster}

\author{L. Casamiquela\inst{1}, R. Carrera\inst{2}, L. Balaguer-Núñez\inst{1}, C. Jordi\inst{1}, C. Chiappini\inst{3}, F. Anders\inst{3}, T. Antoja\inst{1}, N. Miret-Roig\inst{1}, M. Romero-Gómez\inst{1}, S. Blanco-Cuaresma\inst{4}, E. Pancino\inst{5,6}, D. S. Aguado\inst{7,8}, A. del Pino\inst{9}, L. D\'iaz-P\'erez\inst{7,8}, C. Gallart\inst{7,8}}
\institute{Departament de Física Quàntica i Astrofísica, Universitat de Barcelona, ICC/IEEC, 08007 Barcelona, Spain \email{laiacf@fqa.ub.edu} \and
INAF - Osservatorio Astronomico di Padova, Padova, Italy \and
Leibniz-Institut f\"ur Astrophysik Potsdam (AIP), Potsdam, Germany \and
Harvard-Smithsonian Center for Astrophysics, Cambridge, MA 02138, USA \and
INAF - Osservatorio Astrofisico di Arcetri, 50125 Firenze, Italy \and
ASI Science Data Center, 00133 Roma, Italy \and
Instituto de Astrof\'isica de Canarias, La Laguna, 38205 Tenerife, Spain \and
Departamento de Astrof\'isica, Universidad de La Laguna, 38207 Tenerife, Spain \and
Space Telescope Science Institute, Baltimore, MD 21218, USA
}
\authorrunning{L. Casamiquela et al.}
\date{--}

\abstract{The stellar [$\alpha$/Fe] abundance is sometimes used as a proxy for stellar age, following standard chemical evolution models for the Galaxy, as seen by different observational results.}{In this work we show that the Open Cluster NGC6705/M11 has a significant $\alpha$-enhancement [$\alpha$/Fe]$>0.1$ dex, despite its young age ($\sim$300 Myr), challenging the current paradigm.}{We use high resolution (R$>65,000$) high signal-to-noise ($\sim$70) spectra of 8 Red Clump stars, acquired within the OCCASO survey. We determine very accurate chemical abundances of several $\alpha$ elements, using an equivalent width methodology (Si, Ca and Ti), and spectral synthesis fits (Mg and O).}{We obtain [Si/Fe]=$0.13\pm0.05$, [Mg/Fe]=$0.14\pm0.07$, [O/Fe]=$0.17\pm0.07$, [Ca/Fe]=$0.06\pm0.05$ and [Ti/Fe]=$0.03\pm0.03$. Our results place these cluster within the group of young [$\alpha$/Fe]-enhanced field stars recently found by several authors in the literature. The ages of our stars have an uncertainty of around 50 Myr, much more precise than for field stars. By integrating the cluster's orbit in several non-axisymmetric Galactic potentials, we establish the M11's most likely birth radius to lie between 6.8-7.5 kpc from the Galactic center, not far from its current position.}{With the robust Open Cluster age scale, our results prove that a moderate [$\alpha$/Fe]-enhancement is no guarantee for a star to be old, and that not all $\alpha$-enhanced stars can be explained with an evolved blue straggler scenario. Based on our orbit calculations, we further argue against a Galactic bar origin of M11.}
 
\maketitle

\section{Introduction}\label{sec:intro}
The stellar [$\alpha$/Fe] ratio has been widely used as an indirect age estimator because $\alpha$-elements, e.g. O, Mg, Si, Ca, and Ti, are produced in short time scales by core collapse type II supernovae in comparison with iron, synthesized on longer timescales by type Ia supernovae \citep[][among others]{Matteucci2001,Fulbright+2007}. Therefore, as soon as type Ia supernovae, related to intermediate-mass binary systems with mass transfer, start to contribute to the iron enrichment, [$\alpha$/Fe] inevitably decreases. In this scenario, an [$\alpha$/Fe] enhancement means that the star was born in a gas mainly enriched by massive stars.

The correlation between [$\alpha$/Fe] and age has been widely used in the literature to trace the different Galaxy components \citep[e.g.][]{AlvesBrito+2010}. From the analysis of HIPPARCOS stars in a local sphere with a radius of 25 pc \citet{Fuhrmann2011} assigned an age older than 10 Gyr to those stars with high [$\alpha$/Fe] ratios and assumed that they belong to the chemical thick disk. However, recent analysis of larger samples outside the local volume have shown that [$\alpha$/Fe] enhancement does not guarantee that a star is old. \citet{Chiappini+2015} reported the existence of a young [$\alpha$/Fe]-enhanced population from the analysis of a sample of 606 red giants observed by both CoRoT \citep{Miglio+2013} and APOGEE \citep{Majewski+2017}. They showed that most of the stars follow the behaviour of $\alpha$-element abundances predicted by standard evolution models. However, several young stars show unexpectedly high [$\alpha$/Fe] abundances. Interestingly, most of these stars are located at small Galactocentric distances, around 6 kpc from the Galactic center. \citet{Chiappini+2015} also points out the fact that young $\alpha$-enhanced stars are also present in other works available in the literature \citep[e.g.][]{Haywood+2013,Bensby+2014,Bergemann+2014}.  Additionally, at least 14 stars with [$\alpha$/Fe]$>$0.13 and ages younger than 6 Gyr have been detected by \citet{Martig+2015} in their analysis of the stars in common by both \textit{Kepler} and APOGEE, known as APOKASC.

There are different possible scenarios to explain the origin of these young [$\alpha$/Fe] enhanced stars. The first one is a possible ambiguity in determining ages from masses in asteroseismology, since higher masses are assigned to younger ages. As stated by \citet{Jofre+2016} and \citet{Yong+2016}, $\alpha$-rich stars may look young because they have accreted material from a binary companion or because they are a result of a binary merger (blue straggler). In this case, the mass would not reflect the real age of the progenitor star. In the case that these stars are genuinely young, they could have been formed from a recent gas accretion event. Another interpretation is that they could be born in a region near the corrotation of the bar where gas can be kept inert for a long time reflecting only type II supernovae ejecta. Then they could be kicked to their current location. 

In this paper we focus on NGC~6705 (M~11), a young Open Cluster (OC) located in the inner disk $(l,b)=(27.307^{\circ},-2.776^{\circ})$ at a Galactocentric distance of 6.8 kpc and very close to the plane at $z=-90$ pc \citep{Dias+2002}. It has been extensively studied because it is among the most massive known OCs, containing several thousand Solar masses \citep{Santos+2005}. The age of NGC~6705 has been derived from isochrone fitting \citep[e.g.][]{Sung1999,Santos+2005,Beaver+2013,cantatgaudin+2014b} and also from detached eclipsing binaries \citep{Bavarsad+2016}. All of them agree on an age between $\sim$0.2 and 0.3 Gyr. It has been targeted by two of the massive Galactic spectroscopic surveys, APOGEE and GES \citep{Gilmore+2012}, with still controversial results about its chemical composition \citep{cantatgaudin+2014b,Magrini+2014,Tautvaisiene+2015,Magrini+2015,Magrini+2017}.

In the framework of the Open Cluster Chemical Abundances from Spanish Observatories survey \citep[OCCASO][]{Casamiquela+2016} we have obtained very high-resolution (R$\sim$65,000-85,000) high signal-to-noise ratio (SNR$\gtrsim$70) spectra for 8 stars in this cluster. We derived a mild $\alpha$-enhancement which is still outside the expectations of standard chemical evolution models. In this paper we present our chemical analysis of NGC~6705 and we discuss our results according to the chemical evolution models. The paper is organized as follows. In Sect.~\ref{sec:obs.mat} the observational material used is presented. The spectroscopic analysis is detailed in Sect.~\ref{sec:spec}, where we present stellar atmosphere parameters (Sect.~\ref{sec:atmpar}) and chemical abundances (Sect.~\ref{sec:abOCCASO}). An extensive comparison with literature is done in Sect.~\ref{sec:6705literature}. We compute the orbit of the cluster in the Galaxy under different assumptions in Sect.~\ref{sec:6705_orbit}. A discussion of the results is done in Sect.~\ref{sec:discussion}, and the overall conclusions are presented in Sect.~\ref{sec:conclusions}.

\section{Observational material}\label{sec:obs.mat}

OCCASO \citep[see][ Paper I hereafter, for a detailed description]{Casamiquela+2016} is obtaining very high-resolution spectra (R$\gtrsim$65,0000) in the optical range (5000-8000\AA) for Red Clump stars in Northern OCs. This survey is systematically targeting OCs with at least 6 stars per cluster, and with a SNR around 70. It is a natural complement to the GES-UVES observations of OCs from the North, and an optical counterpart for APOGEE.

NGC~6705 has been observed as part of the OCCASO survey. A total of 8 stars have been observed with HERMES \citep{Raskin+2011} installed at Mercator telescope (La Palma, Spain). Three of them were also observed with FIES \citep{Telting+2014} at Nordic Optical Telescope (La Palma, Spain) as part of a subsample designed for comparison between instruments. Details of the observed stars are listed in Table~\ref{tab:stars}: coordinates, magnitude, instrument, SNR of the spectra, atmospheric parameters (see next section) and radial velocities. We also include the available membership information from previous studies: probabilities from proper motion and membership classification from radial velocity.

Radial velocities for the observed stars were presented in Paper I. All observed stars are compatible with being members from their radial velocities within 1$\sigma$. Since then, with new observational runs, we observed one more star (W1256) which has also a compatible radial velocity. The cluster mean radial velocity using the 8 stars is 35$\pm$1 km s$^{-1}$.

\begin{figure}
 \centering
 \includegraphics[width=0.4\textwidth]{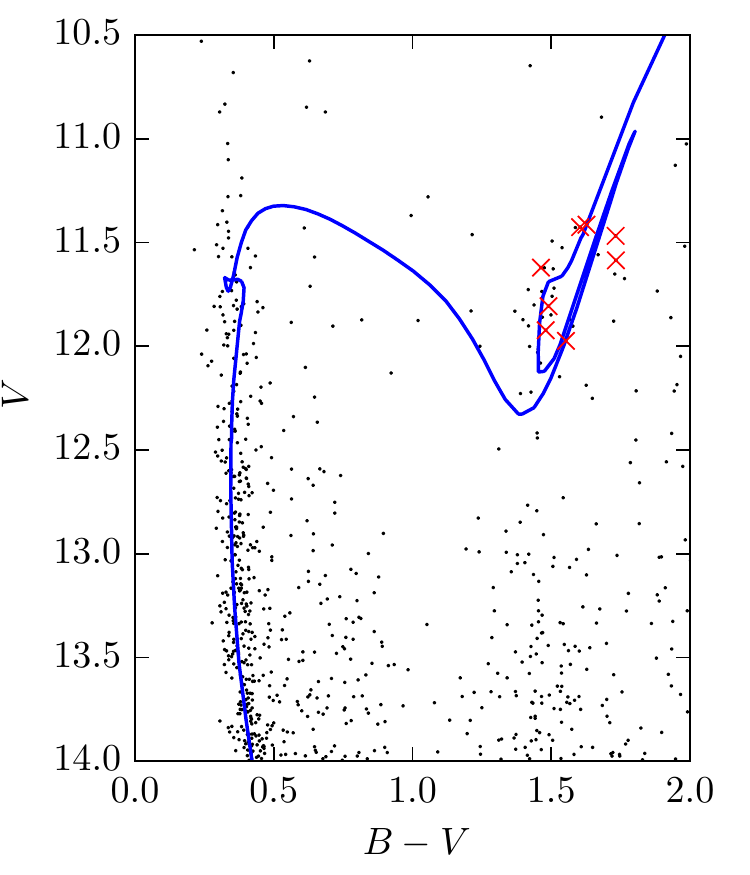}
 \caption{\label{fig:cmd}Colour-magnitude diagram with the photometry from \citet{Sung+1999}. Target stars are marked with red crosses. A PARSEC isochrone of age 316 Myr and $Z=0.019$ shifted by $V-M_{\mathrm{V}}=11.45$ ($d=1950$ pc) and $E\left( B-V \right)=0.40$ \citep{cantatgaudin+2014b} is overplotted.}
\end{figure}

In Fig.~\ref{fig:cmd} we plot the position of the target stars in the colour-magnitude diagram from \citet{Sung+1999}. We overplot a PARSEC isochrone \citep{Bressan+2012} with the age, metallicity, extinction and distance derived by \citet{cantatgaudin+2014b}: age=$316\pm50$ Myr, $Z=0.019$, $E\left( B-V \right)=0.40\pm0.03$ and $V-M_{\mathrm{V}}=11.45\pm0.2$ ($d=1950\pm200$ pc).

\begin{table*}
 \centering
 \caption{\label{tab:stars}Details of the observed stars. Magnitudes and membership data from literature: probability of membership from proper motion, and classification of the star from radial velocity. We list the used instrument and the SNR of the spectra, and the obtained parameters from our spectroscopic analysis. [Fe/H] and $\xi$ are from EW. ID numbers are from WEBDA.}
 {\small
 \setlength\tabcolsep{2.2pt}
 \begin{tabular}{cccccccccccccc}
  \hline
Star ID & RA & DEC & $V$ & $P_{\mathrm{PM}}$ & Class & Instrument & SNR & $T_{\mathrm{eff}}$ & $\log g$ & $\xi$ & $v_{\mathrm{r}}$ & [Fe/H] \\
 & (h:m:s) & (d:m:s) & (mag) & & & & & (K) & (dex) & (km s$^{-1}$) & (km s$^{-1}$) & (dex) \\
\hline
W0660 & 18:51:15.691 &-06:18:14.47 & 11.8 & 0.99 & SM$^a$,SM$^b$,M$^c$& HERMES & 57 & $4738 \pm 53$ & $2.29 \pm 0.13$ & $1.60 \pm 0.09$ & $35.6 \pm 1.0$ & $0.20 \pm 0.05$ \\
W0669 & 18:51:15.318 &-06:18:35.51 & 11.9 & 0.98 & SM$^a$,SM$^b$,M$^c$& HERMES & 54 & $4749 \pm 77$ & $2.23 \pm 0.11$ & $1.72 \pm 0.13$ & $34.5 \pm 1.8$ & $0.21 \pm 0.05$ \\
W0686 & 18:51:14.507 &-06:16:54.74 & 11.9 & 0.99 & SM$^a$,SM$^b$,M$^c$& HERMES & 59 & $4825 \pm 93$ & $2.36 \pm 0.16$ & $1.85 \pm 0.12$ & $36.2 \pm 1.8$ & $0.14 \pm 0.05$ \\
W0779 & 18:51:11.141 &-06:14:33.76 & 11.4 & 0.98 & SM$^a$,M$^c$& HERMES & 65 & $4335 \pm 26$ & $1.70 \pm 0.12$ & $1.45 \pm 0.17$ & $34.3 \pm 1.1$ & $0.19 \pm 0.05$ \\
 & & & & & & FIES & 92 & $4343 \pm 84$ & $1.83 \pm 0.13$ & $1.47 \pm 0.15$ & $34.4 \pm 0.9$ & $0.18 \pm 0.05$ \\
W0916 & 18:51:07.847 &-06:17:11.89 & 11.6 & 0.99 & SM$^a$,SM$^b$& HERMES & 73 & $4789 \pm 67$ & $2.12 \pm 0.27$ & $1.76 \pm 0.13$ & $34.7 \pm 2.2$ & $0.17 \pm 0.05$ \\
W1184 & 18:51:01.989 &-06:17:26.50 & 11.4 & 0.99 & SM$^a$,SM$^b$& HERMES & 70 & $4407 \pm 24$ & $1.76 \pm 0.09$ & $1.34 \pm 0.08$ & $33.1 \pm 0.9$ & $0.13 \pm 0.05$ \\
 & & & & & & FIES & 74 & $4370 \pm 25$ & $1.78 \pm 0.09$ & $1.66 \pm 0.10$ & $33.2 \pm 0.8$ & $0.03 \pm 0.05$ \\
W1256 & 18:51:00.194 &-06:16:59.06 & 11.6 & 0.77 & M$^a$,SM$^b$& HERMES & 85 & $4436 \pm 66$ & $1.83 \pm 0.14$ & $1.59 \pm 0.12$ & $35.7 \pm 0.9$ & $0.07 \pm 0.05$ \\
W1423 & 18:50:55.789 &-06:18:14.26 & 11.4 & 0.99 & SM$^a$,SM$^b$,M$^c$& HERMES & 65 & $4424 \pm 56$ & $1.94 \pm 0.11$ & $1.54 \pm 0.10$ & $36.3 \pm 1.1$ & $0.16 \pm 0.05$ \\
 & & & & & & FIES & 79 & $4524 \pm 107$ & $2.15 \pm 0.09$ & $1.47 \pm 0.11$ & $36.4 \pm 0.8$ & $0.22 \pm 0.05$ \\
\hline
 \end{tabular}}
\begin{flushleft}
{\small Proper motion membership probability from \citet{McNamara+1977}. References for classification (from radial velocity): $^a$\citet{cantatgaudin+2014b}/\citet{Tautvaisiene+2015}, $^b$\citet{Mathieu+1986}, $^c$\citet{Mermilliod+2008} (M: member, SM: single member).}
\end{flushleft}
\end{table*}

\vspace{1cm}
\section{Spectroscopic analysis using OCCASO data}\label{sec:spec}
\subsection{Atmospheric parameters and iron abundance}\label{sec:atmpar}
Stellar atmospheric parameters and iron abundances for stars sampled by OCCASO were obtained by \citet[][Paper II hereafter]{Casamiquela+2017}. Briefly, effective temperature T$_{\mathrm{eff}}$, surface gravity $\log g$, microturbulence $\xi$, and iron abundances were derived using two independent methods: \texttt{DAOSPEC} \citep{cantatgaudin+2014,Stetson+2008} + \texttt{GALA} \citep{Mucciarelli+2013} which uses the equivalent width (EW) methodology, and \texttt{iSpec} \citep{BlancoCuaresma+2014} which uses the spectral synthesis fitting (SS) method. In both cases we adopted the MARCS atmosphere models from \citet{Gustafsson2008} computed assuming 1D-LTE, \citet{Grevesse+2007} solar composition, and the standard $\alpha$-enhancement at low metallicities. We use the version 5 of the master line list from GES \citep{Heiter+2015b} which covers the spectral range $4200 \le \lambda \le 9200$ \AA, and contains atomic information for 35 different chemical species. We refer the reader to Paper II for further details.

\begin{figure}
 \centering
 \includegraphics[width=0.4\textwidth]{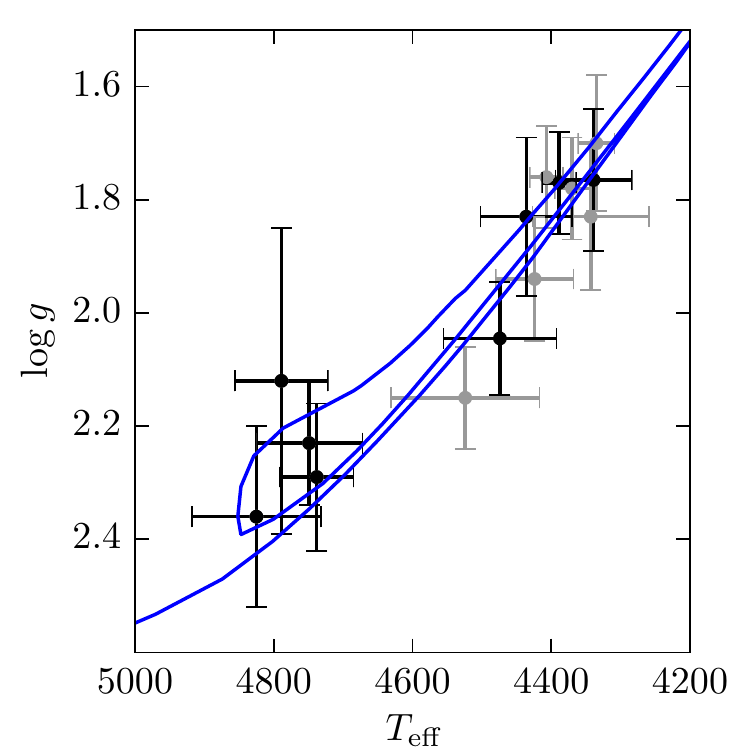}
 \caption{\label{fig:hr}Derived $T_{\mathrm{eff}}$ and $\log g$ in Paper II. Grey points are the values of the stars observed with both FIES and HERMES. For these stars we also plot the mean values (in black). The same isochrone as in Fig.~\ref{fig:cmd} is overplotted.}
\end{figure}

As explained in Paper II, the final stellar atmospheric parameters adopted for each star are the mean of the values obtained by these two methods. Obtained values in the plane T$_{\mathrm{eff}}$-$\log g$ are plotted in Fig.~\ref{fig:hr}, and the same isochrone used in Fig.~\ref{fig:cmd} is drawn as reference. The theoretical position of the red clump traced by this isochrone is well reproduced.

In Paper II, iron abundances were derived also with both methods separately but using the average T$_{\mathrm{eff}}$ and $\log g$ described above. The average iron abundance of the cluster is [Fe/H]$_{\mathrm{EW}}=0.17\pm0.04$ and [Fe/H]$_{\mathrm{SS}}=0.04\pm0.05$, from the EW and the SS analysis, respectively. We refer the reader to Paper II for more details of the two calculations. In the current paper, we use the value derived from EW since it is consistent with other determinations available in the literature.

\subsection{Si, Ca, and Ti chemical abundances}\label{sec:abOCCASO}

The chemical abundances of Si, Ca and Ti were obtained using the EW method.

The EW analysis is performed in two steps. First, EWs are measured using \texttt{DOOp} \citep{cantatgaudin+2014} with \texttt{DAOSPEC} \citep{Stetson+2008}. \texttt{DOOp} \citep{cantatgaudin+2014} is an automatic wrapper for \texttt{DAOSPEC} which finds absorption lines in a stellar spectrum, fits the continuum, measures EWs, identifies lines from a provided line list, and gives a radial velocity estimate. Obtained EWs are fed to \texttt{GALA} \citep{Mucciarelli+2013}, which uses the set of Kurucz abundance calculation codes \citep{Kurucz2005,Sbordone+2004} to derive chemical abundances. Those lines that yield systematically discrepant abundances with respect to the mean abundance of the element are rejected. For this we used the spectra of all stars targeted by OCCASO (115 stars). This is a consistent procedure to discard blended features or lines with inaccurate atomic parameters, obtaining a robust sample of lines for chemical abundance determination from their EW.

The EWs obtained for the 11 analyzed spectra are listed in Table~\ref{tab:EWs}. The uncertainties of the EW measurements are typically below 5\%. For each spectrum we obtained the [X/Fe] ratio as the mean of the abundances derived from individual lines. These values are listed in Table~\ref{tab:abundfin}. The errors in [X/Fe] for each star were estimated from the dispersion of the line abundances divided by the square root of the number of lines, quadratically summed to the error in [Fe/H].

\begin{table*}
 \centering
 \caption{\label{tab:EWs}Lines used to compute abundances of Si, Ca and Ti. We include the atomic information, the reference for the $\log gf$ values, and the derived EWs from \texttt{DAOSPEC}, for all the spectra. The complete table is available online.}
 \footnotesize
 \begin{tabular}{cccccccc}
  \hline
$\lambda$ & Element & $\log gf$ & $\chi$ & EW$_{\mathrm{W0660HER}}$ & EW$_{\mathrm{W0669HER}}$ & ... & EW$_{\mathrm{W1423FIE}}$ \\
(\AA) & & (dex) & (eV) & (m\AA) & (m\AA) & ... & (m\AA) \\
  \hline
5261.704 & \ion{Ca}{I} & -0.579 & 2.52 & 149 & 149 & ... & 168 \\
5349.465 & \ion{Ca}{I} & -0.31  & 2.71 & 150 & 173 & ... & 180 \\
5512.980 & \ion{Ca}{I} & -0.464 & 2.93 & 126 & 127 & ... & 138 \\
  \hline
 \end{tabular}
\end{table*}

\begin{table*}
 \centering
 \caption{\label{tab:abundfin}Final Si, Ca, Ti, Mg and O over Fe abundances of all the analyzed stars. For the stars observed with two instruments we also list average and standard deviation of the two values. The number of used lines is indicated in parentheses.}
 \footnotesize
 \begin{tabular}{lccccc}
  \hline
Spectrum & [Si/Fe] & [Ca/Fe] & [Ti/Fe] & [Mg/Fe] & [O/Fe] \\
  \hline
NGC6705W0660\_HER & $0.05\pm0.05$ (17) & $0.01\pm0.05$ (14) & $-0.03\pm0.04$ (46) & $-0.02\pm0.09$ (2) & $0.08\pm0.06$ \\
NGC6705W0669\_HER & $0.19\pm0.07$ (13) & $0.07\pm0.06$ (14) & $0.03\pm0.05$ (41)  & $0.19\pm0.10$ (2) & $0.12\pm0.07$ \\
NGC6705W0686\_HER & $0.14\pm0.06$ (13) & $0.05\pm0.06$ (14) & $0.03\pm0.05$ (39)  & $0.16\pm0.10$ (2) & - \\
NGC6705W0779\_HER & $0.10\pm0.06$ (17) & $0.10\pm0.06$ (14) & $0.11\pm0.05$ (47)  & $0.08\pm0.09$ (2) & $0.14\pm0.07$ \\
NGC6705W0779\_FIE & $0.12\pm0.06$ (14) & $0.01\pm0.06$ (14) & $-0.02\pm0.05$ (39) & $0.23\pm0.09$ (2) & $0.08\pm0.07$ \\
                  & $0.11\pm0.06$      & $0.06\pm0.08$      & $0.04\pm0.08$       & $0.15\pm0.09$     & $0.11\pm0.06$ \\
NGC6705W0916\_HER & $0.20\pm0.06$ (14) & $0.17\pm0.06$ (13) & $0.06\pm0.05$ (39)  & $0.18\pm0.10$ (2) & $0.26\pm0.07$ \\
NGC6705W1184\_HER & $0.08\pm0.05$ (17) & $0.09\pm0.05$ (14) & $0.12\pm0.05$ (47)  & $0.00\pm0.09$ (2) & $0.23\pm0.06$ \\
NGC6705W1184\_FIE & $0.06\pm0.05$ (14) & $0.05\pm0.05$ (14) & $0.01\pm0.04$ (38)  & $0.23\pm0.09$ (2) & $0.30\pm0.06$ \\
                  & $0.07\pm0.05$      & $0.07\pm0.05$      & $0.06\pm0.07$       & $0.12\pm0.09$     & $0.27\pm0.05$ \\
NGC6705W1256\_HER & $0.07\pm0.05$ (17) & $0.03\pm0.05$ (14) & $0.05\pm0.05$ (47)  & $0.03\pm0.09$ (2) & - \\
NGC6705W1423\_HER & $0.16\pm0.06$ (17) & $-0.02\pm0.05$ (14)& $0.02\pm0.05$ (46)  & $0.18\pm0.09$ (2) & $0.13\pm0.07$ \\
NGC6705W1423\_FIE & $0.08\pm0.05$ (15) & $0.02\pm0.06$ (14) & $0.03\pm0.05$ (38)  & $0.18\pm0.09$ (2) & $0.24\pm0.07$ \\
                  & $0.12\pm0.07$      & $0.00\pm0.06$      & $0.02\pm0.05$       & $0.18\pm0.09$     & $0.19\pm0.10$ \\
  \hline
  Mean$\pm$s.d. &  $0.13 \pm 0.05$ & $0.06 \pm 0.05$ & $0.03 \pm 0.03$ & $0.14 \pm 0.07$ & $0.17 \pm 0.07$ \\
  \hline
 \end{tabular}
\end{table*}

\subsection{O and Mg chemical abundances}\label{sec:MgO}
To determine abundances of O and Mg we performed a SS analysis with \texttt{Salvador} (A. Mucciarelli, priv. comm.). Using this type of analysis we account for blends or hyperfine structure splitting, present in the lines of those elements. \texttt{Salvador} is a tool that fits individual lines from an observed spectrum with synthetic spectra. The spectra are synthesized using the set of Kurucz codes \citep{Kurucz2005,Sbordone+2004} and the MARCS atmosphere models \citep{Gustafsson2008} from the atmospheric parameters determined in Paper II in a window around a given spectral line. The region around the feature of interest is renormalized by the code using the ratio between the observed and the best-fit spectrum. \texttt{Salvador} allows the user to modify the normalization, the size of the window, and the abundances of the different chemical species involved with respect to the assumed by the model (used for blended lines).

Atomic information for the lines are indicated in Table~\ref{tab:MgOlines}.
To determine the Mg abundances we used two lines at 5711.088, and 6318.717 \AA, respectively. Their hyperfine structure splitting was taken into account in the line list. We used the mean of the two lines to derive the overall Mg abundance per spectrum. The mean of the errors determined for the two lines was used as error for [Mg/H], and we quadratically summed the error in [Fe/H] to determine $\sigma$[Mg/Fe]. The abundances for each line and spectra are listed in Table~\ref{tab:MgOab}.

\begin{table*}
 \centering
 \caption{\label{tab:MgOlines}Atomic parameters (oscillator strength and excitation potential) and references for the two Mg lines used, the O forbidden line, and the Ni blend.}
 \footnotesize
 \begin{tabular}{cccccccc}
  \hline
  $\lambda$ & Element & $\log gf$ & $\chi$ & Ref \\
  (\AA) & & (dex) & (eV) &  \\
  \hline
  5711.088 & \ion{Mg}{I} & -1.830 & 4.346 & Kurucz 2010 \\
  6318.717 & \ion{Mg}{I} & -2.020 & 5.108 & Kurucz 2010 \\
  6300.304 & \ion{O}{I}  & -1.830 & 0.0 & \citet{Caffau+2008} \\
  6300.338 & \ion{Ni}{I} & -2.110 & 4.266 & \citet{Johansson+2003} \\
  \hline
 \end{tabular}
\end{table*}

\begin{table*}
 \centering
 \caption{\label{tab:MgOab} Absolute abundances per spectrum of the two Mg lines used, and the O forbidden line.}
 \footnotesize
 \begin{tabular}{cccccccc}
  \hline
  $\lambda$ & Element & $A_{\mathrm{W0660HER}}$ & $A_{\mathrm{W0669HER}}$ & ... & A$_{\mathrm{W1423FIE}}$\\
  (\AA) & & & & ... & \\
  \hline
  5711.088 & \ion{Mg}{I} & $7.85\pm0.10$ & $8.18\pm0.10$ & ... & $7.89\pm0.10$ \\
  6318.717 & \ion{Mg}{I} & $7.68\pm0.07$ & $7.78\pm0.07$ & ... & $7.95\pm0.07$\\
  6300.304 & \ion{O}{I}  & $8.92\pm0.05$ & $9.00\pm0.05$ & ... & $8.92\pm0.05$ \\
  \hline
 \end{tabular}
\end{table*}

For oxygen we use the forbidden [\ion{O}{I}] line at 6300.304~\AA. It is well known that this feature is blended with a \ion{Ni}{I} line \citep{AllendePrieto+2001}, often neglected, but that has a high impact at solar and supersolar metallicities. The code has the option to fix an abundance variation of Ni to perform an accurate fit. So for each spectrum we set this to the Ni abundance derived with the EW methodology typically using around 25 lines. Since oxygen and carbon are bound together by the molecular equilibrium in stellar atmospheres, while determining the oxygen we take into account carbon abundances. We adopt the mean carbon abundance for the cluster to the value [C/H]=-0.08$\pm$0.06 derived by \citet{Tautvaisiene+2015}, to perform the fit with \texttt{Salvador}. The oxygen line could not be measured in W0686 because the large noise prevent us to properly determine the continuum, neither for W1256, which has a sky line on top of it.

To compute the errors on the derived chemical abundances for each line we take into account two sources of uncertainty: the errors due to the choice of the atmospheric parameters, and the errors due to the fit.
\begin{itemize}
 \item To account for the uncertainty due to the assumed atmospheric parameters ($T_{\mathrm{eff}}$, $\log g$ and $\xi$) the abundances are calculated by altering each of them by $\pm1\sigma$. The uncertainty is the standard deviation of the obtained values.
 \item The uncertainty due to the fit is evaluated by performing $N$ MonteCarlo simulations of one line with a desired SNR. In other words, after the fitting procedure, the code takes the best-fit spectrum for that line, adds Poisson noise in order to simulate the provided SNR and repeat the fit; this process is repeated $N$ times. We took $N$=100 and the lowest SNR that we have (54). This procedure accounts for the error due to the SNR and partially to the continuum placement.
\end{itemize}

\begin{figure}
 \centering
 \includegraphics[width=0.5\textwidth]{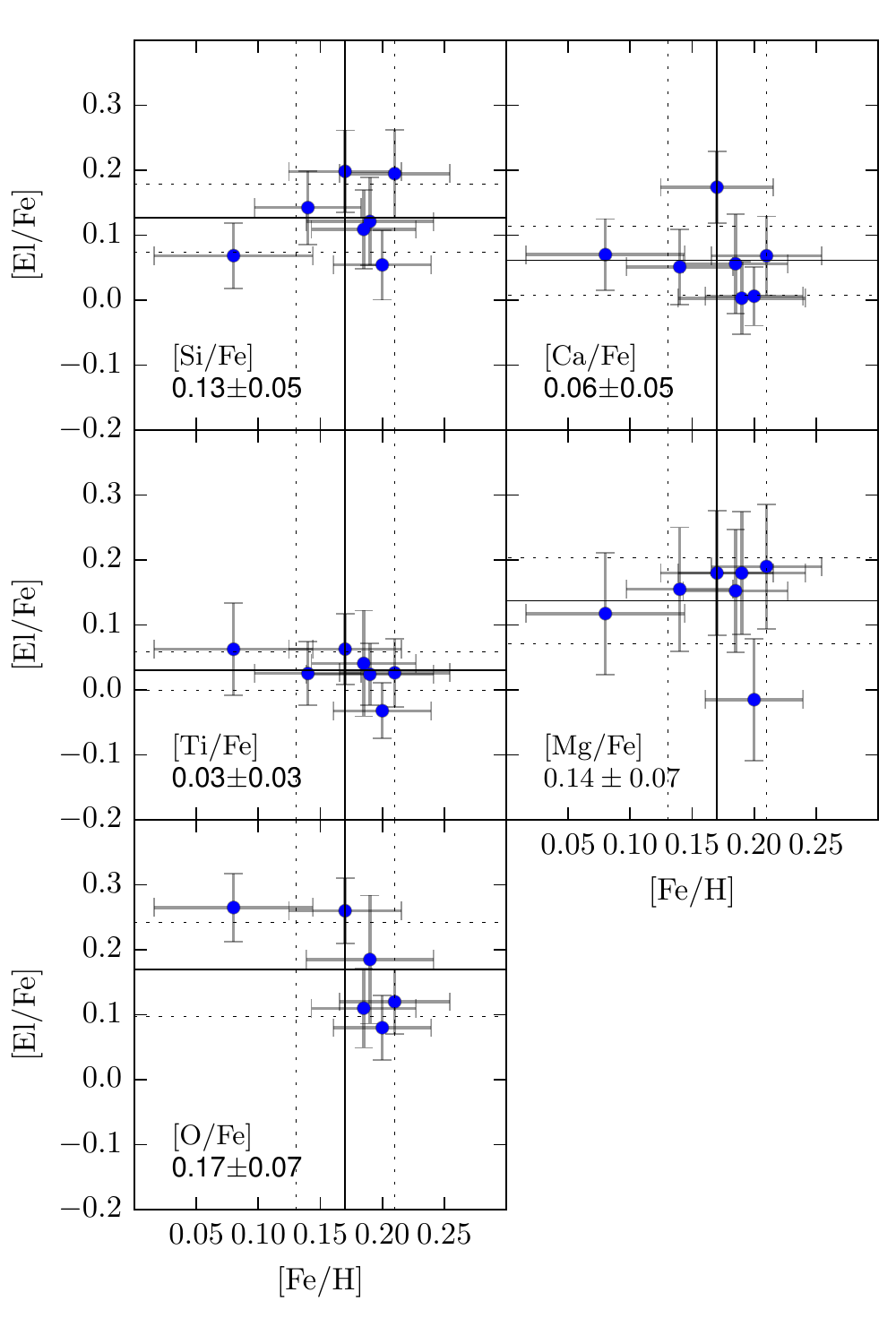}
 \caption{\label{fig:ab_ratios}Abundance ratios of Si, Ca, Ti, Mg and O for the member stars in OCCASO data. Mean abundances (solid lines) and standard deviations (dotted lines) are overplotted in each panel. We have used the 7 member stars except for oxygen, which was calculated with 6 members (see text).}
\end{figure}

The resulting mean [X/Fe] abundances and their spread per spectrum and per star are listed in Table~\ref{tab:abundfin}. The distribution of [X/Fe] vs [Fe/H] per star is plotted in Fig.~\ref{fig:ab_ratios}. We have noticed that the star W1256 has lower abundances (within $1\sigma$) than the rest of the members in all the elements [X/H]. Moreover it has a lower probability of membership from proper motion than the other stars (77\% from Table~\ref{tab:stars}). For these reasons we excluded this star for the very detailed study in this paper, and we derived the average cluster abundances with 7 stars. 

\subsection{Solar abundance scale}
The definition of the Solar abundance scale is key to calculate overall differential abundances with respect to the Sun. To define this scale we have derived Si, Ca, Ti, Mg and O abundances in $16$ Solar spectra available in the Gaia FGK Benchmark Stars (GBS) high resolution spectral library \citep{Blanco+2014}. These spectra have been acquired with different instruments and all of them have been smoothed to match the OCCASO resolution. We use the same line selection and model atmospheres as for the rest of OCCASO stars. We take the atmospheric parameters of the Sun derived in \citet{Heiter+2015}.

The derived mean Solar abundances for each element are: $A\left(\mathrm{Si}\right)_{\odot}=7.48\pm0.01$, $A\left(\mathrm{Ca}\right)_{\odot}=6.25\pm0.02$, $A\left(\mathrm{Ti}\right)_{\odot}=4.91\pm0.02$, $A\left(\mathrm{Mg}\right)_{\odot}=7.58\pm0.01$, $A\left(\mathrm{O}\right)_{\odot}=8.61\pm0.04$. The quoted errors are computed as the standard deviation of the 16 values. These values are consistent within $1-2\sigma$ with previous determinations of the Solar abundance scale, such as \citet{Asplund+2009}: $A\left(\mathrm{Si}\right)_{\odot,A09}=7.51\pm0.01$, $A\left(\mathrm{Ca}\right)_{\odot,A09}=6.29\pm0.02$, $A\left(\mathrm{Ti}\right)_{\odot,A09}=4.91\pm0.03$, $A\left(\mathrm{Mg}\right)_{\odot}=7.53\pm0.01$, $A\left(\mathrm{O}\right)_{\odot,A09}=8.69\pm0.05$.

\section{Comparison with the literature}\label{sec:6705literature}

Chemical abundances from high resolution spectra for NGC~6705 stars have been obtained by several authors including \citet[][GW2000]{Gonzalez+2000}; the several analysis of the GES sample by \citet{cantatgaudin+2014b}, \citet{Magrini+2014}, \citet{Tautvaisiene+2015}, \citet{Magrini+2015}, and \citet{Magrini+2017}; and the latest APOGEE data releases \citep[][]{AlbaretiSDSS+2016,Abolfathi+2017}. In this section we perform a star-by-star comparison for the stars in common with these studies. Finally, we compared the mean abundances derived for NGC~6705 with all these studies.

\citet{Gonzalez+2000} analyzed high‐quality spectra, SNR$\geq$85, for 10 bright K giants in NGC~6705. They used the echelle spectrograph at CTIO 4m telescope (R$\sim$24,000; 4,200-10,000\AA) and the vacuum‐sealed echelle spectrograph at APO 2.5m telescope (R$\sim$34,000; 5,100-8,800\AA). They derived chemical abundances using both SS and EW methods. In the framework of GES, 27 stars have been observed with UVES (R$\sim$47,000; 4,700-7,000\AA). The spectra have been analyzed independently by different teams using different methods \citep{Smiljanic+2014}. Different data releases include additional data and results from different combination of the analysis methods. Finally, high resolution (R=22,500) near infrared (H-band) spectra for several stars in the field of view of NGC~6705 have been obtained by APOGEE \citep{Majewski+2017}. Atmospheric parameters and chemical abundances have been derived using the ASPCAP pipeline \citep{GarciaPerez+2016}. At the moment there is not a dedicated paper to NGC 6705 from APOGEE data, for this reason, we have used the two sets of kinematical and chemical information related through SDSS data releases 13 \citep[DR13][]{AlbaretiSDSS+2016}, and 14 \citep[DR14][]{Abolfathi+2017}. Aside from the new data acquired between the two data releases, there are several changes in the data analysis of DR14 which include a new normalization scheme and a different treatment of the microturbulence \citep[see][for details]{Holtzman+2017}.

\subsection{Star-by-star comparison of stellar parameters} 

\begin{figure}
 \centering
 \includegraphics[width=0.5\textwidth]{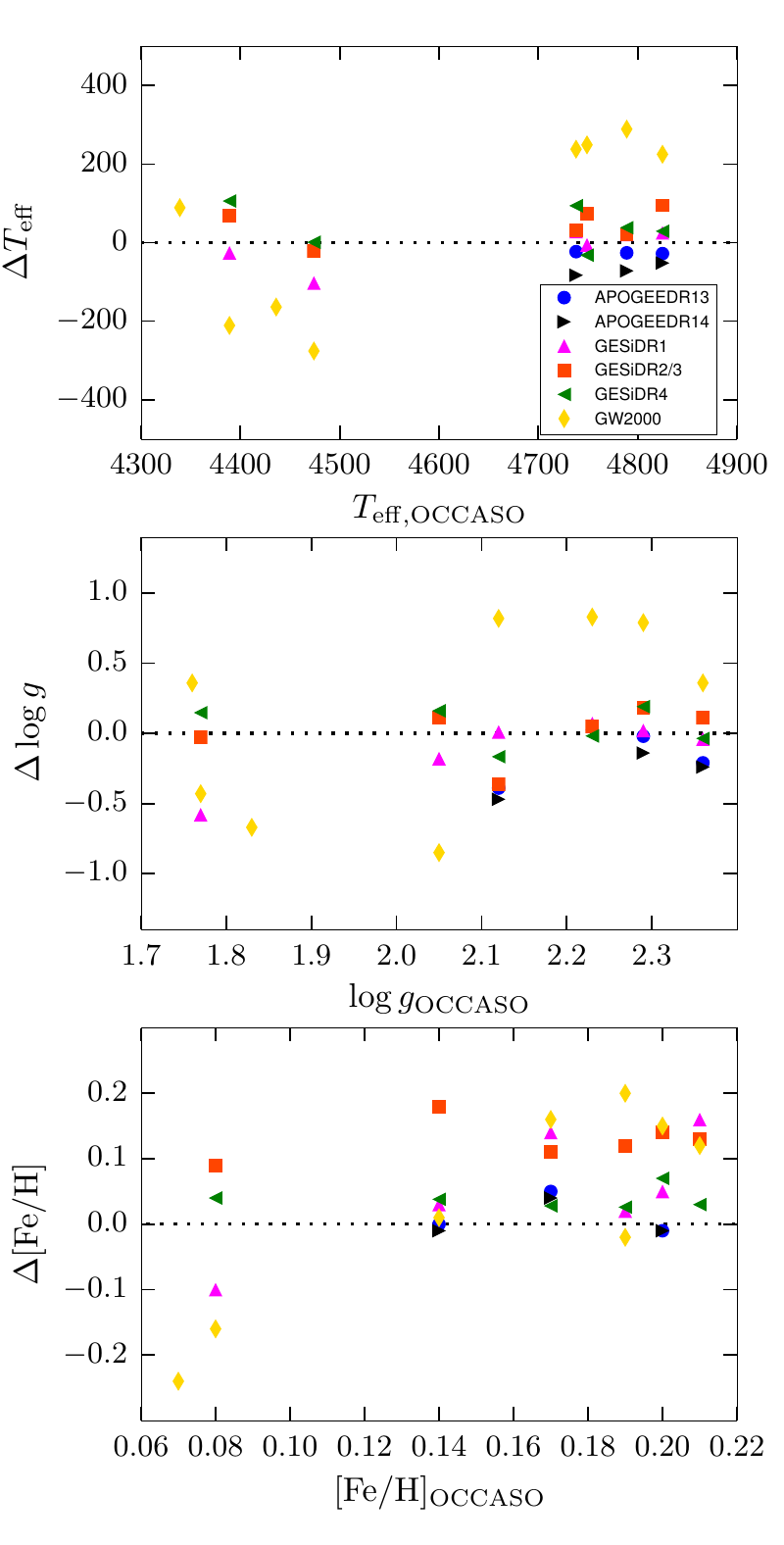}
 \caption{\label{fig:lit}Star-by-star comparison of OCCASO results for $T_{\mathrm{eff}}$, $\log g$ and [Fe/H] with previous high-resolution studies: APOGEE \citet{AlbaretiSDSS+2016}, \citet{Abolfathi+2017}, GES \citep{Magrini+2014,Tautvaisiene+2015,Magrini+2017} and \citet{Gonzalez+2000}. Differences are in the direction OCCASO-literature. Mean differences are listed in Table~\ref{tab:literature}.}
\end{figure}

In total we have 8 stars in common with GW2000, 6 stars with GES and 3 with APOGEE. In Fig.~\ref{fig:lit} we compare the values of T$_{\mathrm{eff}}$ (top), $\log g$ (middle), and [Fe/H] (bottom) derived by the different authors with those obtained here. For each parameter, the mean differences and standard deviations between each sample and OCCASO are listed in Table~\ref{tab:literature} for an exhaustive comparison.

In each data release APOGEE provides two different sets of abundance values namely uncalibrated and calibrated, respectively. Uncalibrated values are those obtained directly by ASPCAP and used to perform the abundance analysis. The calibrated ones are derived using an empirical relation obtained from a sample of well characterized stars in the literature as a function of temperature, gravity and metallicity \citep[see][for details]{Nidever+2015}. This relation changes in the different data releases. The two sets of values are significantly different, specially in $\log g$, for which the uncalibrated ones yield large differences with OCCASO. We limit our comparison to the calibrated parameters.

From the offsets and spreads found comparing OCCASO with the different authors in the literature we can draw some conclusions:
\begin{itemize}
\item In general, $T_{\mathrm{eff}}$ shows mild offsets and dispersions, that agree with the observational errors. However, as can be observed in Fig.~\ref{fig:lit}, GW2000 values show a spread of 219 K, outside their large uncertainties of $\sim150\,$K. Moreover, they show a systematic trend: positive differences are observed for stars with T$_{\mathrm{eff}}>4700\,$K while for cooler stars the differences have opposite sign. 
\item $\log g$ in GESiDR1 is on average 0.12 dex higher than in OCCASO, while in GESiDR2/3 and GESiDR4 show quite a good agreement with the values obtained by OCCASO. APOGEE derives larger surface gravities by 0.21 and 0.28 dex, in DR13 and DR14, respectively. This discrepancy for APOGEE $\log g$ values has already been reported in the literature \citep[e.g.][]{Holtzman+2015}. Finally, GW2000 yields a mean difference of 0.15 dex but again with a large dispersion of 0.65 dex. This may be explained by their large uncertainties in the $\log g$ determination of $\sim0.4$ dex.
\item Except for GESiDR2/3 the [Fe/H] values of all the samples are in good agreement with the OCCASO ones even for GW2000, and in spite of the T$_{\mathrm{eff}}$ and $\log g$ differences discussed above. The GESiDR2/3 values differ also with the other GES samples. The sets that are most similar to OCCASO results and with less dispersion are APOGEE and GESiDR4 ones.
\end{itemize}

\begin{table}
 \centering
 \caption{\label{tab:literature}Mean differences and standard deviations in atmospheric parameters and iron abundances between OCCASO (EW) and literature for the five references that have studied NGC~6705 with high-resolution spectroscopy. Differences are in the direction OCCASO-literature.}
 \setlength\tabcolsep{2.2pt}
 \def\arraystretch{1.2}
 \small
 \begin{tabular}{ccccccccccc}
  \hline
Reference & $\Delta T_{\mathrm{eff}}$ (K) & $\Delta \log g$& $\Delta$[Fe/H] & Num. stars\\
\hline
APOGEEDR13$^1$& $-26\pm2$  &$-0.21\pm0.16$ & $0.01\pm0.03$ & $3$ \\
APOGEEDR14$^2$& $-69\pm13$ &$-0.28\pm0.14$ & $0.00\pm0.02$ & $3$ \\
GESiDR1$^3$   & $-10\pm46$ &$-0.12\pm0.22$ & $0.05\pm0.09$ & $6$ \\
GESiDR2/3$^4$ & $45\pm39$  &$ 0.01\pm0.18$ & $0.13\pm0.03$ & $6$ \\
GESiDR4$^5$   & $39\pm48$  &$ 0.05\pm0.13$ & $0.04\pm0.01$ & $6$ \\
GW2000$^6$    & $55\pm219$ &$ 0.15\pm0.65$ & $0.03\pm0.13$ & $8$ \\
  \hline
 \end{tabular}
 \begin{flushleft}
 $^1$\citet{AlbaretiSDSS+2016}, $^2$\citet{Abolfathi+2017}, $^3$\citet{Magrini+2014}, $^4$\citet{Tautvaisiene+2015}, $^5$\citet{Magrini+2017}, $^6$\citet{Gonzalez+2000}.
 \end{flushleft}
\end{table}

\subsection{Cluster mean abundances from literature}\label{sec:OCablit} 

In this section we compare our mean abundances with those derived by the studies presented above. For all these comparisons we use the mean values and uncertainties obtained by each of them.

In the case of APOGEE, we retrieve stars in a radius of $16$ arcmin around the cluster center and with radial velocities in the range $32<v_{\mathrm{r}}<38$ km s$^{-1}$. This range has been selected from the analysis of OCCASO radial velocities in Paper I. We have rejected those stars with radial velocity uncertainty larger than 0.5 km s$^{-1}$ since they are potential spectroscopic binaries \citep{Nidever+2015}. Since APOGEE Ti abundances are less reliable, we have excluded this element from our comparison \citep[see][for a detailed discussion]{Hawkins+2016b}. We have also excluded stars with Fe, Si, Ca, Mg and O abundances discrepant with respect of the bulk of the cluster. We obtain a total of 11 potential member stars in both data releases.

In Table~\ref{tab:abratios+lit} we list cluster mean abundance determinations by the cited studies in comparison with OCCASO results:

\begin{itemize}
\item There is a very good agreement between the abundances obtained here and those derived by GW2000 in spite of the large uncertainties involved in their analysis and the large differences in temperature and gravity discussed in the previous section.
\item In the case of APOGEE, there is a good agreement for [Fe/H] values. [Si/Fe] and [Ca/Fe] are slightly lower in APOGEE but the differences are well within the uncertainties. However, there is a difference in Mg of $\sim$0.20 dex, and in O of 0.18 and 0.28 dex with APOGEE DR13 and DR14, respectively. We note that APOGEE retrieves O abundances from molecules, mainly CO and OH, while we used the forbidden [\ion{O}{I}] atomic line.
\item In the case of GES, [Fe/H] in iDR1 and iDR4 are slightly lower than in the case of OCCASO but still within the uncertainties, but in iDR2/3 it is lower by 0.17 dex. In general, the values obtained by GES are always lower than those derived by OCCASO, except for [Mg/Fe] in iDR1 and iDR4, and [O/Fe] in iDR2/3, which is an interactive SS determination by \citet{Tautvaisiene+2015}.
\end{itemize}

Summarizing, GW2000 shows an [$\alpha$/Fe] enhancement similar to those derived in OCCASO. The same is true for APOGEE if we excluded Mg and O from the analysis. This behavior is not observed in the case of GES except for [Mg/Fe] and [O/Fe] in iDR2/3 \citep{Tautvaisiene+2015,Magrini+2015}. \citet{Magrini+2015}, based on 27 stars, reported a high mean [$\alpha$/Fe] with respect to chemical evolution models. They conclude that it is genuine, and they explore the possibility that this cluster has suffered from the effect of a local enrichment by a supernova type II.

\begin{table*}
 \centering
 \caption{\label{tab:abratios+lit}Mean iron and $\alpha$-element abundance calculated in this study: using OCCASO results (using $6$ or $7$ member stars, depending on the chemical species), and using APOGEE DR13 and DR14 results from \citet[][$11$ member stars]{AlbaretiSDSS+2016,Abolfathi+2017}. We include the results obtained from the three GES data releases GESiDR1 \citep[][21 member stars]{Magrini+2014}, GESiDR2/3 \citep[][27 members]{Tautvaisiene+2015,Magrini+2015}, and GESiDR4 \citep[][15 members]{Magrini+2017}.}
 \setlength\tabcolsep{2pt}
 \def\arraystretch{1.1}
 \begin{tabular}{lccccccc}
  \hline
 Element & OCCASO & GW2000 & \multicolumn{2}{c}{APOGEE} & \multicolumn{3}{c}{GES} \\
 & & & DR13 & DR14 & iDR1 & iDR2/3 & iDR4 \\
  \hline
  $[$Fe/H$]$  & $0.17\pm0.04$ & $0.17\pm0.13$ & $0.16\pm0.03$ & $0.16\pm0.03$  & $0.14\pm0.06 $& $0.00\pm0.05$ & $ 0.12\pm0.05$\\
  $[$Si/Fe$]$ & $0.13\pm0.05$ & $0.14\pm0.16$ & $0.10\pm0.04$ & $0.09\pm0.03$ & $0.03\pm0.05 $& $>0.10$ & $ 0.02\pm0.07$\\
  $[$Ca/Fe$]$ & $0.06\pm0.05$ & $0.10\pm0.14$ & $-0.04\pm0.02$ & $-0.05\pm0.03$ & $-0.02\pm0.05$& $-$ & $-0.07\pm0.09$\\
  $[$Ti/Fe$]$ & $0.03\pm0.03$ & $0.05\pm0.06$ & $-$ & $-$  & $-0.05\pm0.07$& $-$           & $-0.04\pm0.09$\\
  $[$Mg/Fe$]$ & $0.14\pm0.07$ & $-$ & $-0.08\pm0.03$ & $-0.09\pm0.03$  & $0.20\pm0.09 $& $>0.10$           & $ 0.10\pm0.07$\\
  $[$O/Fe$]$  & $0.17\pm0.07$ & $-$ & $0.01\pm0.03$ & $-0.09\pm0.04 $ & $          - $& $0.13\pm0.05$ & $-0.13\pm0.07$\\
  \hline
 \end{tabular}
\end{table*}

\section{Orbit computation}\label{sec:6705_orbit}
We studied if the peculiarity in $\alpha$ element abundances shown by this OC could be partly explained by a very different birth and current Galactocentric radii \citep{Sellwood+2002} (i.e. born in the inner Galaxy near the bar and then migrated outwards, see Sect.~\ref{sec:intro}). To check that, we reconstructed the orbit of the cluster and integrated it backwards until the time of birth. To do so one basically needs: age, 3D position ($l$, $b$, and heliocentric distance $d$), 3D velocity (proper motions and radial velocity $\mu_{\alpha}\cos \delta$, $\mu_{\delta}$, $v_{\mathrm{r}}$), and to assume a certain gravitational potential for the Milky Way\footnote{Other parameters are needed like Sun position and velocity: $R_0=8.34\,$kpc, $(U,V,W)_0=(10.7,15.6,8.9)\,$km s$^{-1}$ and Galactic rotation $240\,$km s$^{-1}$ \citep{Reid+2014}.}. We used a gravitational potential that includes Galactic bar and spiral arms resembling those of the Milky Way: a prolate bar from \citet{Pichardo+2004}, and the spiral arms from the PERLAS model in \citet{Pichardo+2003}. We refer to the cited references for details of the model and the parameters that best fit the Milky Way.

This method can carry large uncertainties: (i) errors coming from the assumed distances, motions and age; (ii) inaccuracies on the assumed model of the gravitational potential, (e.g. axisymmetric, featuring the bar and/or spiral arms), and the free parameters involved in them; (iii) for the old OCs the assumption of a static potential is not a correct approximation taking into account that typical pattern speeds of the dynamic structures can change in few Gyr.
Since NGC~6705 OC is young, the uncertainties that come from assuming a static potential when integrating back the orbit are small.

In this section we examine in detail the propagation of errors in the assumed motions, distance, and age. We also quantify the uncertainties that come from the choice of the model.

We use three sets of proper motions and distances specified in Table~\ref{tab:dataX_6705} to compute the orbits. Data1 uses the mean of $8$ stars from TGAS data (Cantat-Gaudin et al. 2017, submitted), data2 uses proper motions and distances from \citet{Dias+2002}, data3 uses the mean of 32 stars (Casamiquela et al. 2017, submitted) from TGAS. In all cases we adopt $v_{\mathrm{r}}=34.5\pm1.7\,\mathrm{km\,s}^{-1}$ (result from Paper I), and the age $316\pm50$ Myr \citep[derived by][]{cantatgaudin+2014b}. For each dataset we sweep $91$ different model parameters of the gravitational potential. We have explored values of: spiral arms mass ($0$, $0.03$, $0.05$ in units of disc mass, $8.56\times10^{10}\,M_{\odot}$), spiral arms pattern speed ($15$, $20$, $30$ km s$^{-1}$ kpc$^{-1}$), mass of the bar ($0$, $0.6$, $0.8$ in units of bulge mass, $1.41\times10^{10}\,M_{\odot}$), bar pattern speed ($36$, $46$, $56$ km s$^{-1}$ kpc$^{-1}$) and bar orientation respect to the Sun ($20$, $40$ deg). For each model and set of input parameters we do $100$ realizations of the orbit integration assuming Gaussian errors in the motions, distance and age.

\begin{table}[h]
 \centering
 \caption{The three datasets of distances and proper motions, assumed in the computation of the birth radius. We include the median birth radius and its standard deviation from $91$ models.}\label{tab:dataX_6705}
 \setlength\tabcolsep{2pt}
 \def\arraystretch{1.1}
 \begin{tabular}{cccccccc}
  \hline
 Reference & $d$ & $\mu_{\alpha}\cos \delta$ & $\mu_{\delta}$ & $R_{\mathrm{GC,birth}}$ \\
 & (kpc) & ($\mathrm{mas\,yr}^{-1}$) & ($\mathrm{mas\,yr}^{-1}$) & (kpc) \\
 \hline
 data1$^i$ & $1.754$ &$-1.93 \pm0.39$     & $-4.88\pm 0.42$ & $7.0 \pm0.1$ \\
 data2$^{ii}$ & $1.877$ &$-1.23 \pm3.85$  & $ 1.31\pm 4.32$ & $7.9 \pm0.1$ \\
 data3$^{iii}$ & $1.647$ &$-1.04 \pm0.25$ & $-3.80\pm 0.30$ & $7.4 \pm0.1$ \\
 \hline
 \end{tabular}
 
 \flushleft
 Reference for distances and proper motions: $^i$ from TGAS, mean of 8 stars (Cantat-Gaudin et al. 2017, submitted), $^{ii}$ \citet{Dias+2002}, $^{iii}$ from TGAS, mean of 32 stars (Casamiquela et al. 2017, submitted). 
 \end{table}

\begin{figure}
 \centering
 \includegraphics[width=0.5\textwidth]{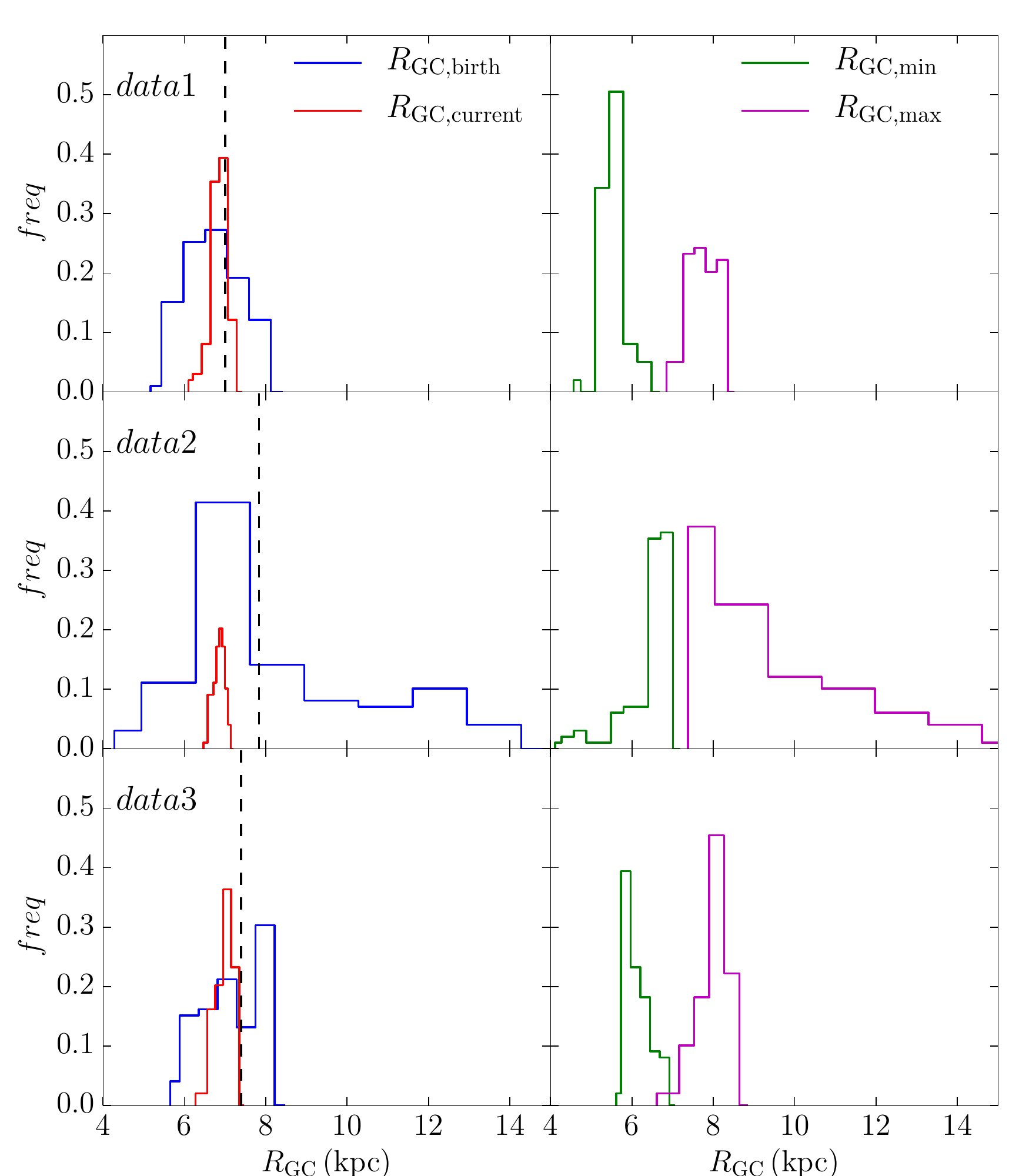}
 \caption{Distributions of current and birth radii (left), and the maximum and minimum orbit radius (right), given by $100$ realizations of the one of the models of the gravitational potential (see text). Each row shows the results for the three datasets specified in Table~\ref{tab:dataX_6705}. The median birth radius is indicated with a dashed vertical line.}\label{fig:orbit6705_model32}
\end{figure}

\begin{figure}
 \centering
 \includegraphics[width=0.5\textwidth]{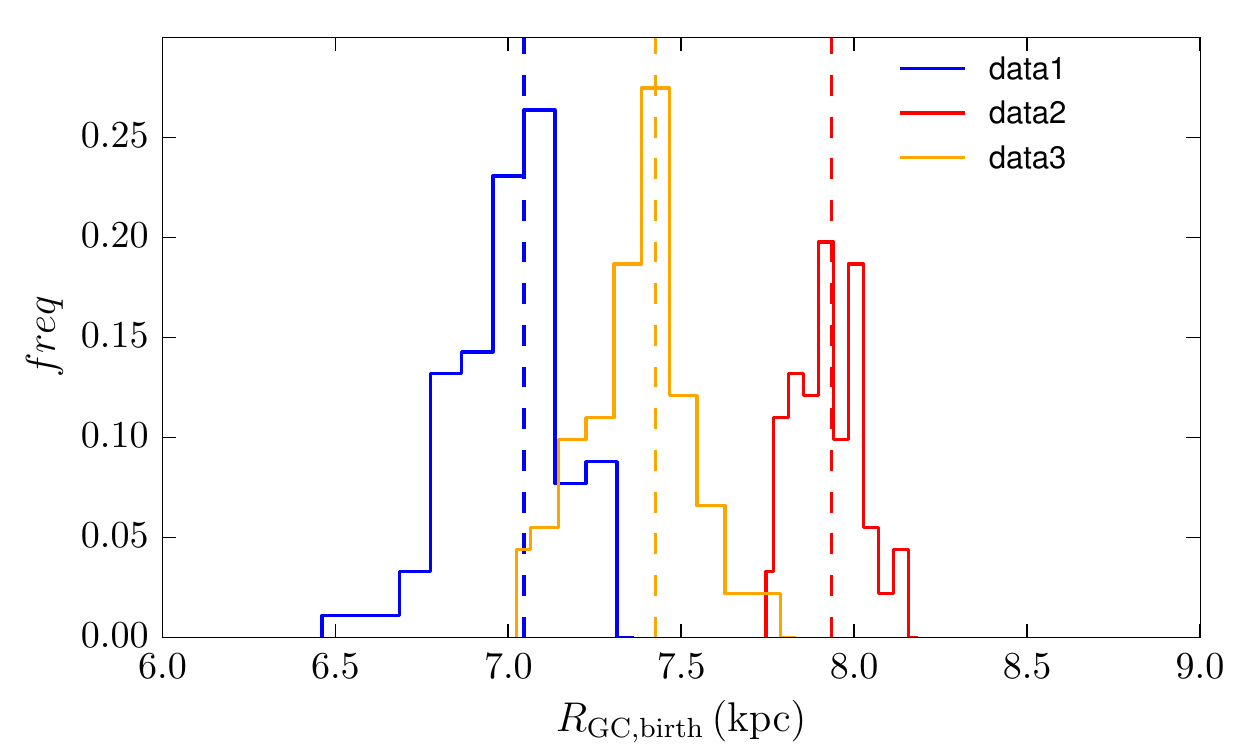}
 \caption{Normalized distribution of the birth radius of the orbits given by $91$ different models of the gravitational potential, and the three datasets of assumed observational parameters in Table~\ref{tab:dataX_6705}. The median values of birth radius are plotted as vertical dashed lines.}\label{fig:Rbirth_dist}
\end{figure}

The results are the following:
\begin{enumerate}
 \item Each run of a model has a distribution of possible orbits given the assumed errors. An example for one of the models\footnote{Spiral arms mass of $0.03$ units of disc mass, spiral pattern speed of $20$ km s$^{-1}$ kpc$^{-1}$, mass of the bar of $0.6$ units of bulge mass, bar pattern speed of $46$ km s$^{-1}$ kpc$^{-1}$, and bar orientation of $20$ deg.} is plotted in Fig.~\ref{fig:orbit6705_model32}, where we show the distribution of the current and birth radii, the minimum and maximum radius of the orbits. Here, the median and standard deviation of birth radius are $7.0\pm0.7$, $7.8\pm2.5$, $7.4\pm0.8$ kpc for data1, data2 and data3, respectively. In the three cases this model predicts that the cluster was born slightly outwards or almost at the current radius. The current radii computed for each datasets are very similar: $6.9\pm0.2$, $6.9\pm0.1$, $7.1\pm0.2$ kpc for data1, data2 and data3, respectively. Results of data1 and data3 are very similar, since the source of distance and proper motions is the same, though different membership selection make the birth radius differ by $0.2$ kpc. Data2 results of birth, minimum and maximum radii are quite different, and they also show a larger spread and longer tails, because the errors are larger.
 \item The distribution of birth radius given by the $91$ models and the $3$ datasets are plotted in Fig.~\ref{fig:Rbirth_dist}. It is seen that the three determinations of proper motions and distance lead to a significant difference in the computed orbits, and in particular in the radius at birth. We list the mean values of $R_{\mathrm{GC,birth}}$ in Table~\ref{tab:dataX_6705}. In the case of data1, in $66$\% of the models the radius at birth is between $6.9<R_{\mathrm{GC,birth}}<7.2$ kpc ($1\sigma$ from the median). For data2 $60$\% of the cases lie within $1\sigma$, $7.8<R_{\mathrm{GC,birth}}<8.0$ kpc. In the case of data3 we obtain $7.3<R_{\mathrm{GC,birth}}<7.6$ kpc in $63$\% of the models.
\end{enumerate}

From those results, which take into account different models of the gravitational potential, different sources of the data and errors in the proper motions, radial velocity, distance and age, we can conclude that the Galactocentric radius at birth of NGC~6705 is between $6.8<R_{\mathrm{GC,birth}}<8.9$ kpc. Since we have lower errors for data1 and data3, we can say that with high probability it would be between $5.4<R_{\mathrm{GC,birth}}<7.5$ kpc, slightly inside the Solar radius. Taking into account all the models and error realizations the birth radius is lower than $5$ kpc only in $0.98$\%, $1.40$\%, $0.13$\% of the cases for data1, data2, and data3, respectively.

\section{Discussion}\label{sec:discussion}
From the results of OCCASO data in Sect.~\ref{sec:spec} we see that this cluster shows a clear overabundance in $\alpha$ elements: Si, Mg and O show an enhancement of $0.13\pm0.05$, $0.14\pm0.07$ and $0.17\pm0.07$ respectively. Ca and Ti show more moderate values of $0.06\pm0.05$ and $0.03\pm0.03$, respectively. From the comparison with other high-resolution studies (Sect.~\ref{sec:6705literature}, Table~\ref{tab:abratios+lit}): APOGEE finds similar values of [Fe/H] and [Si/Fe] as in OCCASO; in the different GES data releases a similar enhancement is derived in Si (iDR1), Mg and O (iDR2/3); and GW2000 finds mainly the same values of [Fe/H], [Si/Fe], [Ca/Fe] and [Ti/Fe]. Our resolution is much higher ($R\sim65,000-85,000$) than in all previous studies, and this plays a major role when analyzing spectra of metal-rich stars. On the other hand, as discussed by \citet{Magrini+2015} for the GES results of this cluster, the $\alpha$-enhancement in Si and Mg is genuine and not due to NLTE effects.

The literature is contradictory in the abundances of the different elements, but our results point towards a young metal-rich and $\alpha$-enhanced OC, at least in some of the $\alpha$-elements.
Indeed, different enhancement levels for different $\alpha$ elements are expected on nucleosynthetic grounds, because the contribution of thermonuclear SN Type Ia to Si and Ca is expected to be more pronounced than for O or Mg \citep[at least in standard SNIa models, e.g.][, with small amounts of unburned material]{Iwamoto+1999}. This is exactly what our measurements seems to imply, with [O/Fe] > [Mg/Fe] > [Si/Fe] or [Ca/Fe], thus suggesting a protocluster cloud polluted with a larger fraction of  type II SNe/type Ia SNe than the typical local gas. 

As stated in Sect.~\ref{sec:intro} this feature is seen in some samples of field stars.
Unlike the stars analysed by \citet{Chiappini+2015}, where the age has been inferred from the measurement of the mass (via the combination of seismic and spectroscopic information), and hence could be affected by later mass accretion or even merger in case of binary stars, the cluster age is much more robust. In the latter case, there is no doubt the cluster is young, and its age is determined with unquestionable precision and accuracy, with an uncertainty of 50 Myr. This is hence a genuine young-alpha rich object, and could give some support to the claim of \citet{Chiappini+2015} and \citet{Martig+2015} that some of their stars could be indeed young, and that the evolved blue-straggler scenario \citep{Jofre+2016,Yong+2016} does not explain the whole picture. In addition, the fact that NGC 6705 is located in the thin disk also supports the idea that the young $\alpha$ rich stars found in CoRoT inner-disc field could be thin disk stars despite their current position above the plane (z$\sim$-300pc).

\citet{Chiappini+2015} argued, by computing the guiding radii of their analyzed stars, that they find a preferential location towards the inner Galaxy, thus giving support to the idea that these stars have a common origin near the Galactic bar. 
From the results in Sect.~\ref{sec:6705_orbit} the cluster was probably born roughly inside the Solar radius, but far from the bar.

On the other hand, \citet{Magrini+2015} tries to explain this peculiarity comparing this cluster with the high-$\alpha$ metal-rich population (HAMR) found first by \citet{Adibekyan+2011} and later confirmed by several authors. Comparing with simulations it is plausible for HAMR stars to be extreme migrators born in the inner bulge (R$<2\,$kpc) and that have migrated towards the solar neighbourhood \citep{Adibekyan+2013}. In comparison with this cluster, the age of the HAMR stars is in general much older (in mean older than thin disc stars by about 3 Gyr).

Again, after our orbit analysis for NGC~6705 it seems unlikely that it has a common origin as the HAMR stars, so to be born in the very inner Galaxy. Of course, there are other effects that we are not taking into account in this type of analysis, such as diffusion by giant molecular clouds, transient spiral arms \citep{Roskar+2012} or resonance coupling between bar and spiral arms \citep{Minchev+2010}, that can make the cluster migrate. But the young age of this cluster makes this option highly improbable.

There are other explanations that could apply in this case, such as a local self-enrichment by SN type II in a giant molecular cloud proposed by \citet{Magrini+2015}. Under their calculations an enrichment of more than 0.1 dex in [Mg/Fe], [O/Fe] and [Si/Fe] could be reached due to a SN type II explosion in the mass range $15-18\,\mathrm{M}_{\odot}$.

\section{Conclusions}\label{sec:conclusions}
We have performed an abundance analysis of Fe, Si, Ca, Ti, Mg and O from high-resolution spectroscopic data. First, we have derived abundances from 7 (most likely) members of NGC~6705 from OCCASO spectra. We compare results with those of APOGEE DR13 and DR14 for Ca, Si, Mg and O. Finally, a comparison is shown among OCCASO and the different data releases of GES and APOGEE.

According to OCCASO results this OC is metal rich ([Fe/H]=0.17$\pm$0.04, Paper II) and it shows a clear $\alpha$-enhancement in Si, Mg and O, and a mild enhancement above the errors in Ca and Ti. The mean [Fe/H] abundance found in OCCASO agrees very well with APOGEE, GESiDR1 and GESiDR4, it does not agree with GESiDR2/3.  The mean [Si/Fe] within errors agrees with the APOGEE results and is higher than in the GES; [Ca/Fe] is higher in this work than in APOGEE and GES; [Ti/Fe] within errors agrees with GES, no results are analyzed from APOGEE. The mean [Mg/Fe] agrees well within errors with GES and is higher than the results from APOGEE; the mean [O/Fe] agrees well with the precise GESiDR2 analysis result by \citet{Tautvaisiene+2015}, and is higher than all other results.

We have done a kinematic study of this cluster integrating its orbit back in time to derive its birth place. We use three sets of proper motions and distances, 91 assumptions for the gravitational potential free parameters, and we do 100 realizations of each orbit assuming Gaussian errors in radial velocity, proper motion, distance and age. We conclude that the birth Galactocentric radius of NGC~6705 is probably between $6.8<R_{\mathrm{GC,birth}}<7.5$ kpc. After our analysis it seems unlikely for NGC~6705 to be born near the bar, so its origin is probably different from that of the young $\alpha$-enhanced stars of \citet{Chiappini+2015}. Another possibility could be a local self-enrichment of the giant molecular cloud by a type II SN. A more quantitative investigation should be done to justify the latter case, so the explanation for the genuine $\alpha$-enhancement of this cluster is still open.

\begin{acknowledgements}
We thank G. Tautvai\v siene for refereeing this paper and for her suggestions that improved this work.

This work is based on observations made with the Nordic Optical Telescope, operated by the Nordic Optical Telescope Scientific Association, and the Mercator Telescope, operated on the island of La Palma by the Flemish Community, both at the Observatorio del Roque de los Muchachos, La Palma, Spain, of the Instituto de Astrof\'isica de Canarias. This work is also based on observations collected at the Centro Astron\'omico Hispano Alem\'an (CAHA) at Calar Alto, operated jointly by the Max-Planck Institut f\"ur Astronomie and the Instituto de Astrof\'isica de Andaluc\'ia (CSIC).

This research made use of the WEBDA database, operated at the Department of Theoretical Physics and Astrophysics of the Masaryk University, the SIMBAD database, operated at the CDS, Strasbourg, France, and the TOPCAT tool version 4.4. This work has made use of the VALD database, operated at Uppsala University, the Institute of Astronomy RAS in Moscow, and the University of Vienna. This work was supported by the MINECO (Spanish Ministry of Economy) - FEDER through grant ESP2016-80079-C2-1-R and ESP2014-55996-C2-1-R and MDM-2014-0369 of ICCUB (Unidad de Excelencia 'Mar\'ia de Maeztu'). 

LC acknowledges financial support from the University of Barcelona under the APIF grant.

\end{acknowledgements}

\bibliographystyle{aa}
\bibliography{biblio_v3.bib}

\end{document}